\documentclass[preprintnumbers,amsmath,amssymb,twocolumn]{revtex4}
\usepackage{dcolumn}
\usepackage{graphicx}
\usepackage{subfig}
\usepackage{verbatim}
\usepackage{color} 
\usepackage{amsmath}
\usepackage{array}
\usepackage{multirow}

\begin{document}

\title{Free Energy of Multiple Overlapping  Chains}
\author{Katherine Klymko}
\author{Angelo Cacciuto}
\email{ac2822@columbia.edu}
\affiliation{Department of Chemistry, Columbia University \\New York, New York 10027}
\date{\today}
\begin{abstract}
How accurate is pair additivity in describing interactions between soft polymer-based nanoparticles?
Using numerical simulations we compute the free energy cost required to overlap multiple 
chains in the same region of space, and provide a quantitative measure of the effectiveness of pair additivity as a function of  chain number and length. Our data suggest that pair additivity can indeed become quite inadequate as the chain density in the overlapping region increases.
We also show that even a scaling theory based on polymer confinement can only partially account for the complexity of the problem. In fact, we unveil and characterize an isotropic to star-polymer cross-over taking place for large number of chains, and propose a revised scaling theory that better captures the physics of the problem.
\end{abstract}  
 
\maketitle
The question of how  nano and mesoscopic particles spontaneously organize into complex macroscopic structures
can be considered as one of the great challenges in the field of soft matter today. In fact, the prospect of developing materials with
new and exciting optical, mechanical, and electronic properties via the process of self-assembly relies on our 
ability to predict and control the phase behavior of complex fluids.

Although most of the work on self-assembly has historically focused on small molecules, the last decade has witnessed several breakthroughs in particle synthesis at the meso-scale~\cite{DeVries,Schnablegger,Hong,Weller,Hobbie} making possible the production of  nanoparticles with complex chemical and geometrical properties with an unprecedented degree of precision. 
More recently, a burst of research activities has focused on polymer-based nanoparticles~\cite{LikosRev}. 
What makes these particles very interesting is that they break the dogma of mutual non-penetrability.
Unlike regular  colloids, for which excluded volume interactions are strictly enforced via a hard-core or a power-law potential,  complex mesoparticles such as charged or neutral star polymers, dendrimers or microgels present a very peculiar pair potential describing their volume interactions. Beyond the small and local deformation limit,  which can be easily described in terms of classical elasticity theory, these soft polymer-based nanoparticles have  very unusual interaction potentials which allow for even complete overlap among themselves~\cite{grosberg00,louis,LikosRev}.

Surprisingly, the simple relaxation of excluded volume constraints results 
in an extremely rich phase behavior that has been reported  in several 
publications~\cite{louis,likos0,likosA,likosB,denton,gottawald,pierleoni, Capone, bozorgui,cacciuto2}. 
Notably, it was found that some classes  of soft interactions lead 
to  reentrant melting transitions, others to polymorphic cluster phases~\cite{mladek2}, and in general to multiple  transitions 
involving close-packed and non-close-packed crystalline phases~\cite{suto,LikosNat,pephertz} as a function of the system density.  
Remarkably,  the phase behavior of these systems is very much dependent on the 
the shape of the pair potential, and it is today feasible to engineer interactions between star polymers or dendrimers
by controlling their overall chemical/topological properties~\cite{mladek}. 
Likos et al.~\cite{likos}  established a criterion 
to predict whether for a bounded and repulsive potential re-entrant melting or cluster phases will occur based
on the sign of the Fourier transform of the interaction. For a recent review on the subject we refer the reader to reference~\cite{LikosRev}. 
  
Given the complexity of these nanoparticles,  their interactions are usually extracted via 
an explicit coarse-grained procedure to reduce the problem to a simpler one consisting of single effective particles interacting via an ad-hoc pair potential. 
Once a pair potential as a function of separation $r$, $F(2,r)$   is extracted, pair additivity among any two effective particles 
is assumed and phase diagrams  are computed. 

While it is by now clear that many body effects can lead to density-dependent interactions between 
polymer pairs~\cite{Louis22}, and methods to systematically include such deviations have recently been put forward~\cite{capone22,Marina} (and references therein), in this paper we show that the very assumption of 
pair additivity can be greatly inadequate in describing the interactions among polymer-based nanoparticles.
To prove it, we compute the total free energy cost associated with overlapping $n_p$ self-avoiding polymers (effective particles) of length $N$ and show that very quickly the assumed additivity of the pair interactions breaks down.

If one assumes straightforward  pair additivity, the free energy cost required to overlap $n_p$ effective particles is simply proportional to the number of pair interactions
\begin{equation} 
\beta  F(n_p)=\beta F(2) \frac{n_p(n_p-1)}{2}\label{str}
\end{equation}
  
Alternatively, we can describe $F(n_p)$ by using the scaling theory based on polymer confinement as was first suggested 
by Jun et. al.~\cite{Jun1}. The main idea is that in the dilute limit overlapping two chains of length $N$, each having a radius of gyration 
$R_{\rm G}\sim N^{\nu}$, is equivalent to confining a single chain of length $2N$ into a spherical cavity having a radius equal to 
the radius of gyration of a single chain $R_{\rm G}$. The free energy cost associated with spherical confinement of a polymer of length $N$ into a spherical cavity  of radius $R$ is given by $\beta F\sim  \left(R_{\rm G}/R \right)^{3/(3\nu-1)}$ 
(where $\beta=(k_{\rm B}T)^{-1}$)~\cite{grossberg,cacciuto,sakaue}. By plugging in the denominator $R\rightarrow R_{\rm G}$ and in the numerator $R_{\rm G}\rightarrow (2N)^{\nu}$, one obtains  $\beta F(2)\sim \, 2^{3\nu/(3\nu-1)}$. This equation, easily generalizable to $n_p$ chains  as $\beta F(n_p)\sim \, n_p^{3\nu/(3\nu-1)}$, states that (a) the free energy cost to completely overlap two chains is independent of their length (well established result both theoretically and numerically~\cite{grosberg00,louis}), and (b) that overlapping $n_p$ chains has a free energy cost that does not grow linearly with the number of pair interactions.

This  scaling theory,  that should better account for the density increase  within the overlapping region of the chains, can be 
re-written as
\begin{equation} 
\beta  F(n_p)=\beta F(2)\left (1+\frac{(n_p-2)}{2}\right )^{3\nu/(3\nu-1)} \label{rev}
\end{equation}
For chains in a good solvent $\nu\approx 3/5$, and the value of the exponent is close to $2.25$ and is expected to cross over 
to $3$ for large number of polymers~\cite{Jun1,cacciuto}. 
To establish the correct dependence of the free energy with the number of chains we performed numerical simulations.      

We modeled a chain as a sequence of $N$ spherical beads (monomers) of diameter $\sigma$ connected sequentially with 
an entropic spring of maximum extension $\sqrt 2\sigma$. The interaction between any two monomers is described via a hard-core potential~\cite{FrenkelBook} while the entropic spring between consecutive monomers has the form
\begin{equation} 
\beta  V_s(r)=\begin{cases} 0 \,\,\,\,\,\,\,\,\,\,\,   {\rm if \,\,} \sigma<r<\sqrt2\sigma&\\ \infty  \,\,\,\,\,\,\,\,  {\rm otherwise}\end{cases}
\end{equation} 
The advantage of this model is that the free energy associated with multiple overlapping chains is fully contained in their
configurational entropy.
To compute the free energy  we used the thermodynamic integration method~\cite{FrenkelBook}.
The idea is to introduce a fictitious potential 
\begin{equation} 
\beta  V^{i\neq j}_{\lambda}(r)=\lambda\begin{cases} 0 \,\,\,\,\,\,\,\,\,\,\,\,\,\,\,\,\,\,\,\,\,\,\,\,   {\rm if \,\,} r>\sigma&\\ (1-\frac{r}{\sigma})  \,\,\,\,\,\,\,\,  
{\rm if \,} r\leq\sigma\end{cases}
\end{equation}
which acts exclusively among monomers associated to different chains ($i\neq j$), and to constrain the center of mass of each chain
to be within a spherical shell of radius $r_0=2\sigma$. For $\lambda=0$ the overlapping chains are not interacting;  as 
$\lambda\rightarrow\infty$ we recover the system of interest. The free energy due to the polymer-polymer interactions 
can then be extracted by performing the following integral 
\begin{equation}
 F(n_p)=\int_0^{\infty}d\lambda \left(\frac{d V_{\lambda}}{d\lambda}\right)_{\lambda}
 \end{equation}
 In practice, we perform several  Monte Carlo simulations for numerous values of $\lambda$ until the hard sphere limit is effectively reached, and compute numerically the integral.
In all our data the largest value of $\lambda$ was selected to be the one for which the total energy of the system was 
decreased to a value of the order of $10^{-3}$. 
The small tail of the integral for larger values of lambda was computed  by fitting the tail of the 
data with a power law and by extending the integral until the change of $F(n_p)$ becomes clearly negligible (this typically accounts for 
a very small amount of the total free energy).
To properly sample $V_{\lambda}$ our simulations were run from a minimum of 50 million to a maximum of 250 million sweeps.
To improve our statistics, after every system sweep a global chain rotation move is also implemented. This consists of 
randomly picking a monomer and a direction in a random chain, and of performing a rigid rotation around a randomly selected axis 
of a small random angle of all the monomers connecting the selected monomer to the end of the 
chain along the selected direction. 
Our simulation were performed in the $NVT$ ensemble in the dilute limit ($V\gg 4/3\pi R_{\rm G}^3$) for polymers of length $N=64$ and $N=128$, with $n_p$ ranging from 2 to 32. 

\begin{figure}
	\includegraphics[width=0.38\textwidth]{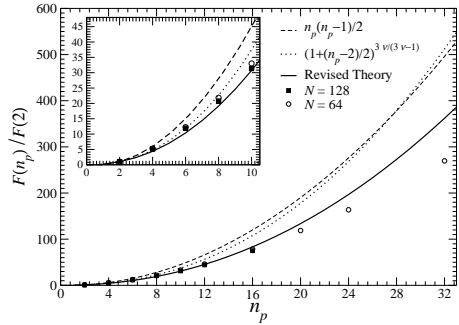}
	\caption{Free energy cost to overlap $n_p$ chains as a function of the number of chains. The dashed and dotted lines represent respectively the straightforward and the scaling-based theory from Eq.~\ref{str} and Eq.~\ref{rev} respectively. The solid line shows 
	the revised theory in Eq.~\ref{newt}. The inset is a zoom of the data for small $n_p$.}
	 \label{free_energy}
\end{figure}

Our results indicate that the free energy required to superimpose two chains equals $\beta F(2)=2.20(5)$ for $N=128$ and
$\beta F(2)=2.31(5)$ for $N=64$. These results are fully consistent with previous numerical simulations 
on similar systems, and the small difference is simply due to finite size effects. Figure~\ref{free_energy} presents the core results of our simulations and shows how $F(n_p)$ normalized 
by the pair free energy $F(2)$ varies with the number of polymers. 
Lines indicating the straightforward  (Eq.~\ref{str}) and  the scaling-based (Eq.~\ref{rev}) predictions are also shown as a reference. Our findings clearly show that Eq.~\ref{str} becomes   inaccurate as soon as $n_p> 4$, Eq~\ref{rev} is accurate up to
$n_p\leq 8$ polymers,  yet significant and systematic deviations from both predictions 
are apparent for larger values of $n_p$. Interestingly, the dependence of the free energy with the number of chains becomes weaker as
the number of chains increases. 

\begin{figure}
	\includegraphics[width=0.38\textwidth]{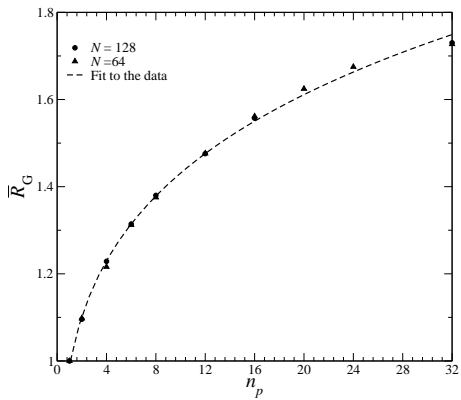}
	\caption{Average radius of gyration $\bar{R}_{\rm G}$ of multiple overlapping chains as a function 
	of the number of chains $n_p$. The line is a fit to the data.}\label{asymmetry}
\end{figure}

\begin{figure}
	\includegraphics[width=0.38\textwidth]{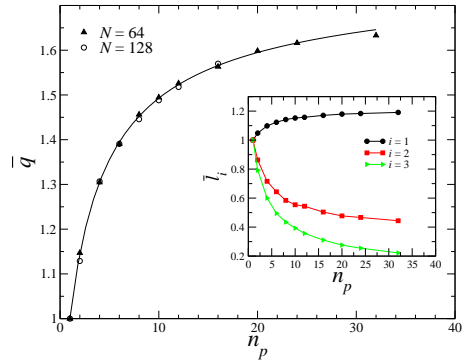}
	\caption{
	Average asphericity  $\bar{q}$ for $N=64$ and $N=128$, computed according to Eq.~\ref{asph}, for multiple overlapping chains as a function 
	of the number of chains $n_p$. The inset shows the three normalized eigenvalues of the shape tensor $\bar{l}_i$ for $N=64$.}\label{qs}
\end{figure}

This result is a bit counter intuitive because one should expect a stronger dependence of $F(n_p)$ on $n_p$ as 
the polymer concentration is increased inside the overlapping area. The key issue with the scaling theory is that 
it assumes that the size of the confining cavity (which equals the radius of gyration of an unconstrained polymer $R_{\rm G}$) remains constant  for any number of polymers. Although this is a good assumption for few chains, we find that  
as $n_p$ increases the average size of the confining cavity becomes systematically larger. This is 
because chains do swell to minimize the number of interactions.  Figure~\ref{asymmetry} shows the average
size of the system, measured by computing the average radius of gyration of the chains at large values of $\lambda$, 
$\langle R_{\rm G}(n_p)|_{\lambda=\infty}\rangle$, divided by the radius of gyration of the reference non-interacting  polymers, 
$\langle R_{\rm G}(n_p)|_{\lambda=0}\rangle$. 
By fitting these data to a power law 
$\bar{R}_{\rm G}(n_p)=a+b\, n_p^{\alpha}$, we can explicitly account for this correction.
The revised free energy scales as 
\begin{equation}
\beta F_{\rm R}(n_p)\sim \left ( n_p^{\nu}/(a+b\,n_p^{\alpha})\right )^{3/(3\nu-1)} \label{newt}
\end{equation}
 with $a\simeq 0.3$,  $b\simeq 0.7$ and $\alpha\simeq 0.21$. As seen in Fig.~\ref{free_energy}, this function successfully  fits most of our numerical data up to $n_p=16$. 
Crucially, the scaling behavior of the radius of gyration with the number of chains $\alpha\approx 1/5$ is consistent with that of 
a star polymer in the swollen regime ($n_p^{1/2} \ll N$). This suggests that a  rearrangement 
from isotropically mixing chains to demixed/localized chains (star-polymer) may be taking place.
To investigate this scenario, we computed the asphericity of the polymers as a function of $n_p$.
This is obtained by computing the inertia tensor of each polymer, and by combining the three eigenvalues 
$l_1$, $l_2$ and $l_3$ into the rotational invariant parameter ~\cite{rudnick}
\begin{equation}
q=\frac{ (l_1-l_2)^2+(l_1-l_3)^2+(l_2-l_3)^2}{2(l_1+l_2+l_3)^2}\label{asph}
\end{equation}
The plot of the normalized
asphericity $\bar q=q(n_p)/q(1)$  as well as the three normalized eigenvalues $\bar {l}_i=l_i(n_p)/l_i(1)$ as a function of the 
number of polymers $n_p$ is shown in Fig.~\ref{qs}. 
For a polymer in a good solvent a first order epsilon expansion results, in the dilute limit  and for $N\rightarrow\infty$,
to a value of $q$ close to 0.415. Accurate numerical simulations predicted a value closer to 0.431~\cite{Jado,Cannon}.
For our longest chains, we find that indeed $q=0.44(2)$ when not interacting, however, as soon as polymers interact, $q$ increases significantly to more than 1.6 times its original. This is a clear indication that the chains are not swelling isotropically as $n_p$ increases, but are stretching out along their long axis at expenses of the other two directions. This morphological change of the chains is suggestive of the fact that chain segregation may also be occurring.

\begin{figure}
	\includegraphics[width=0.43\textwidth]{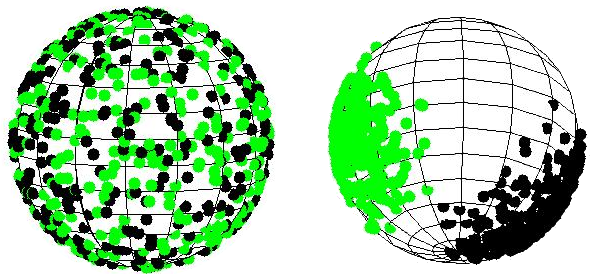}
	\caption{  Chains's axial maps. The l.h.s tracks the location of the main axis of two chains  projected onto a spherical surface centered around the system over the course of a long Monte Carlo trajectory when only two chains are present. 
	The r.h.s shows the same map for two randomly selected chains in a system containing 32 overlapping polymers.  	In the former case the polymers' axes perform a random walk over the spherical surface, indicating complete mixing of the chains, whereas in the latter case the chains are clearly segregated in specific regions.}\label{probs}
\end{figure}

 To check for chain segregation  we tracked the location of the chains' main axes over the course of a long simulation trajectory, and projected it over the surface of a sphere centered around the system. Figure ~\ref{probs} tracks down the locations of polymers' axes for $n_p=2$ and $n_p=32$. For the sake of clarity we show the tracks of only two randomly selected chains also for the case of $n_p=32$. Clearly, when $n_p=2$ the chain's axes can explore the whole spherical surface, indicating complete mixing of the chains. In the latter case only a small region of the surface is explored by the two  selected chains, clear indication of chain segregation.

All of our data points to the fact that our system is effectively behaving as a star-polymer,  i.e. a
system of polymers radiating from a central core, and segregated into roughly conical regions.
A power law fit of our data including only values of the free energy for $n_p\geq 15$ results in
$\beta F(n_p)\propto n_p^{1.6(1)}$ which is indeed compatible with star polymers in the semi-dilute limit~\cite{cotton}.
 
In summary, we computed the free energy cost associated with the complete overlap of multiple self-avoiding chains as a function of chain number. Our data show that although the free energy error associated with pair additivity of potentials between soft polymer-based nanoparticles is of the order of just a few $k_{\rm B}T$   for $n_p\leq 4$ (making the straightforward approach in this limit, if not accurate, a least reasonable), such an approximation severely  deteriorates for larger values of $n_p$. 
 We have also shown that an extended scaling theory based on polymer confinement of a single chain into a spherical cavity 
can better account for the free energy cost of multiple overlapping polymers.
Finally, we have shown that  de-mixing and chain segregation occurs when considering large numbers of chains, and that in this limit,
the free energy of a star-polymer  provides a better description of our numerical data.  

Although our study focuses on single polymer chains, the  main message should also hold for multi-polymer-based particles  
such as star-polymers and dendrimers, where deviations could be much more significant even for small $n_p$ and could
have dramatic consequences especially for the phase diagrams associated with systems forming cluster crystals where multiple particles overlap on the same lattice sites. It should be stressed that our results can be directly linked to the chain segregation process described by Jun et. al.~\cite{Jun1} for polymers under confinement. In this analogy, the formation of a star-polymer  can be understood as chain segregation with the added constraint of overlapping  center of masses.
\section*{ACKNOWLEDGMENTS}
This work was supported by the National Science Foundation under CAREER Grant No. DMR-0846426. 

\begin{thebibliography}{35}
\expandafter\ifx\csname natexlab\endcsname\relax\def\natexlab#1{#1}\fi
\expandafter\ifx\csname bibnamefont\endcsname\relax
  \def\bibnamefont#1{#1}\fi
\expandafter\ifx\csname bibfnamefont\endcsname\relax
  \def\bibfnamefont#1{#1}\fi
\expandafter\ifx\csname citenamefont\endcsname\relax
  \def\citenamefont#1{#1}\fi
\expandafter\ifx\csname url\endcsname\relax
  \def\url#1{\texttt{#1}}\fi
\expandafter\ifx\csname urlprefix\endcsname\relax\def\urlprefix{URL }\fi
\providecommand{\bibinfo}[2]{#2}
\providecommand{\eprint}[2][]{\url{#2}}

\bibitem[{\citenamefont{DeVries et~al.}(2007)}]{DeVries}
\bibinfo{author}{\bibnamefont{DeVries}} \bibnamefont{et~al.},
  \bibinfo{journal}{Science} \textbf{\bibinfo{volume}{315}},
  \bibinfo{pages}{358} (\bibinfo{year}{2007}).

\bibitem[{\citenamefont{Li et~al.}(1999)\citenamefont{Li, Schnablegger, and
  Mann}}]{Schnablegger}
\bibinfo{author}{\bibfnamefont{M.}~\bibnamefont{Li}},
  \bibinfo{author}{\bibfnamefont{H.}~\bibnamefont{Schnablegger}},
  \bibnamefont{and} \bibinfo{author}{\bibfnamefont{S.}~\bibnamefont{Mann}},
  \bibinfo{journal}{Nature} \textbf{\bibinfo{volume}{402}},
  \bibinfo{pages}{393} (\bibinfo{year}{1999}).

\bibitem[{\citenamefont{Hong et~al.}(2006)\citenamefont{Hong, Jiang, and
  Granick}}]{Hong}
\bibinfo{author}{\bibfnamefont{L.}~\bibnamefont{Hong}},
  \bibinfo{author}{\bibfnamefont{S.}~\bibnamefont{Jiang}}, \bibnamefont{and}
  \bibinfo{author}{\bibfnamefont{S.}~\bibnamefont{Granick}},
  \bibinfo{journal}{Langmuir} \textbf{\bibinfo{volume}{22}},
  \bibinfo{pages}{9495} (\bibinfo{year}{2006}).

\bibitem[{\citenamefont{Weller}(2003)}]{Weller}
\bibinfo{author}{\bibfnamefont{H.}~\bibnamefont{Weller}},
  \bibinfo{journal}{Phil. Trans. R. Soc. A} \textbf{\bibinfo{volume}{361}},
  \bibinfo{pages}{229} (\bibinfo{year}{2003}).

\bibitem[{\citenamefont{Hobbie et~al.}(2005)}]{Hobbie}
\bibinfo{author}{\bibfnamefont{E.~K.} \bibnamefont{Hobbie}}
  \bibnamefont{et~al.}, \bibinfo{journal}{Langmuir}
  \textbf{\bibinfo{volume}{21}}, \bibinfo{pages}{10284} (\bibinfo{year}{2005}).

\bibitem[{\citenamefont{Likos}(2006{\natexlab{a}})}]{LikosRev}
\bibinfo{author}{\bibfnamefont{C.~N.} \bibnamefont{Likos}},
  \bibinfo{journal}{Soft Matter} \textbf{\bibinfo{volume}{2}},
  \bibinfo{pages}{478} (\bibinfo{year}{2006}{\natexlab{a}}).

\bibitem[{\citenamefont{Grosberg et~al.}(1982)\citenamefont{Grosberg, Khalatur,
  and Khokhlov}}]{grosberg00}
\bibinfo{author}{\bibfnamefont{A.~Y.} \bibnamefont{Grosberg}},
  \bibinfo{author}{\bibfnamefont{P.~G.} \bibnamefont{Khalatur}},
  \bibnamefont{and} \bibinfo{author}{\bibfnamefont{A.~R.}
  \bibnamefont{Khokhlov}}, \bibinfo{journal}{Makromol. Chem. Rapid Commun.}
  \textbf{\bibinfo{volume}{3}}, \bibinfo{pages}{709} (\bibinfo{year}{1982}).

\bibitem[{\citenamefont{Louis et~al.}(2000)\citenamefont{Louis, Bolhuis,
  Hansen, and Meijer}}]{louis}
\bibinfo{author}{\bibfnamefont{A.~A.} \bibnamefont{Louis}},
  \bibinfo{author}{\bibfnamefont{P.~G.} \bibnamefont{Bolhuis}},
  \bibinfo{author}{\bibfnamefont{J.~P.} \bibnamefont{Hansen}},
  \bibnamefont{and} \bibinfo{author}{\bibfnamefont{E.~J.}
  \bibnamefont{Meijer}}, \bibinfo{journal}{Phys. Rev. Lett.}
  \textbf{\bibinfo{volume}{85}}, \bibinfo{pages}{2522} (\bibinfo{year}{2000}).

\bibitem[{\citenamefont{G\"otze et~al.}(2004)\citenamefont{G\"otze, Harreis,
  and Likos}}]{likos0}
\bibinfo{author}{\bibfnamefont{O.}~\bibnamefont{G\"otze}},
  \bibinfo{author}{\bibfnamefont{H.~M.} \bibnamefont{Harreis}},
  \bibnamefont{and} \bibinfo{author}{\bibfnamefont{C.~N.} \bibnamefont{Likos}},
  \bibinfo{journal}{J. Chem. Phys.} \textbf{\bibinfo{volume}{120}},
  \bibinfo{pages}{7761} (\bibinfo{year}{2004}).

\bibitem[{\citenamefont{N.Likos et~al.}(1998)\citenamefont{N.Likos, Watzlawek,
  Abbas, Jucknischke, Allgaier, and Richter}}]{likosA}
\bibinfo{author}{\bibfnamefont{C.}~\bibnamefont{N.Likos}},
  \bibinfo{author}{\bibfnamefont{H.~L.} \bibnamefont{Watzlawek}},
  \bibinfo{author}{\bibfnamefont{B.}~\bibnamefont{Abbas}},
  \bibinfo{author}{\bibfnamefont{O.}~\bibnamefont{Jucknischke}},
  \bibinfo{author}{\bibfnamefont{J.}~\bibnamefont{Allgaier}}, \bibnamefont{and}
  \bibinfo{author}{\bibfnamefont{D.}~\bibnamefont{Richter}},
  \bibinfo{journal}{Phys. Rev. Lett.} \textbf{\bibinfo{volume}{80}},
  \bibinfo{pages}{4450} (\bibinfo{year}{1998}).

\bibitem[{\citenamefont{Jusufi et~al.}(2002)\citenamefont{Jusufi, Likos, and
  L\"owen}}]{likosB}
\bibinfo{author}{\bibfnamefont{A.}~\bibnamefont{Jusufi}},
  \bibinfo{author}{\bibfnamefont{C.~N.} \bibnamefont{Likos}}, \bibnamefont{and}
  \bibinfo{author}{\bibfnamefont{H.}~\bibnamefont{L\"owen}},
  \bibinfo{journal}{Phys. Rev. Lett.} \textbf{\bibinfo{volume}{88}},
  \bibinfo{pages}{018301} (\bibinfo{year}{2002}).

\bibitem[{\citenamefont{Denton}(2003)}]{denton}
\bibinfo{author}{\bibfnamefont{A.~R.} \bibnamefont{Denton}},
  \bibinfo{journal}{Phys. Rev. E} \textbf{\bibinfo{volume}{67}},
  \bibinfo{pages}{11804} (\bibinfo{year}{2003}).

\bibitem[{\citenamefont{Gottwald et~al.}(2004)\citenamefont{Gottwald, Likos,
  Kahl, and L\"owen}}]{gottawald}
\bibinfo{author}{\bibfnamefont{D.}~\bibnamefont{Gottwald}},
  \bibinfo{author}{\bibfnamefont{C.~N.} \bibnamefont{Likos}},
  \bibinfo{author}{\bibfnamefont{G.}~\bibnamefont{Kahl}}, \bibnamefont{and}
  \bibinfo{author}{\bibfnamefont{H.}~\bibnamefont{L\"owen}},
  \bibinfo{journal}{Phys. Rev. Lett.} \textbf{\bibinfo{volume}{92}},
  \bibinfo{pages}{68301} (\bibinfo{year}{2004}).

\bibitem[{\citenamefont{Pierleoni et~al.}(2006)\citenamefont{Pierleoni,
  Addison, Hansen, and Krakoviack}}]{pierleoni}
\bibinfo{author}{\bibfnamefont{C.}~\bibnamefont{Pierleoni}},
  \bibinfo{author}{\bibfnamefont{C.}~\bibnamefont{Addison}},
  \bibinfo{author}{\bibfnamefont{J.~P.} \bibnamefont{Hansen}},
  \bibnamefont{and}
  \bibinfo{author}{\bibfnamefont{V.}~\bibnamefont{Krakoviack}},
  \bibinfo{journal}{Phys. Rev. Lett.} \textbf{\bibinfo{volume}{96}},
  \bibinfo{pages}{128302} (\bibinfo{year}{2006}).

\bibitem[{\citenamefont{Capone et~al.}(2008)}]{Capone}
\bibinfo{author}{\bibfnamefont{B.}~\bibnamefont{Capone}} \bibnamefont{et~al.},
  \bibinfo{journal}{J. Phys. Chem. B} \textbf{\bibinfo{volume}{113}},
  \bibinfo{pages}{3629} (\bibinfo{year}{2008}).

\bibitem[{\citenamefont{Bozorgui et~al.}(2010)\citenamefont{Bozorgui, Sen,
  Miller, P\'amies, and Cacciuto}}]{bozorgui}
\bibinfo{author}{\bibfnamefont{B.}~\bibnamefont{Bozorgui}},
  \bibinfo{author}{\bibfnamefont{M.}~\bibnamefont{Sen}},
  \bibinfo{author}{\bibfnamefont{W.~L.} \bibnamefont{Miller}},
  \bibinfo{author}{\bibfnamefont{J.~C.} \bibnamefont{P\'amies}},
  \bibnamefont{and} \bibinfo{author}{\bibfnamefont{A.}~\bibnamefont{Cacciuto}},
  \bibinfo{journal}{J. Chem. Phys.} \textbf{\bibinfo{volume}{132}},
  \bibinfo{pages}{014901} (\bibinfo{year}{2010}).

\bibitem[{\citenamefont{Cacciuto and Luijten}(2006)}]{cacciuto2}
\bibinfo{author}{\bibfnamefont{A.}~\bibnamefont{Cacciuto}} \bibnamefont{and}
  \bibinfo{author}{\bibfnamefont{E.}~\bibnamefont{Luijten}},
  \bibinfo{journal}{Nanoletters} \textbf{\bibinfo{volume}{6}},
  \bibinfo{pages}{901} (\bibinfo{year}{2006}).

\bibitem[{\citenamefont{Mladek et~al.}(2006)\citenamefont{Mladek, Gottwald,
  Kahl, Neumann, and Likos}}]{mladek2}
\bibinfo{author}{\bibfnamefont{B.~M.} \bibnamefont{Mladek}},
  \bibinfo{author}{\bibfnamefont{D.}~\bibnamefont{Gottwald}},
  \bibinfo{author}{\bibfnamefont{G.}~\bibnamefont{Kahl}},
  \bibinfo{author}{\bibfnamefont{M.}~\bibnamefont{Neumann}}, \bibnamefont{and}
  \bibinfo{author}{\bibfnamefont{C.~N.} \bibnamefont{Likos}},
  \bibinfo{journal}{Phys. Rev. Lett.} \textbf{\bibinfo{volume}{96}},
  \bibinfo{pages}{045701} (\bibinfo{year}{2006}).

\bibitem[{\citenamefont{Suto}(2006)}]{suto}
\bibinfo{author}{\bibfnamefont{A.}~\bibnamefont{Suto}}, \bibinfo{journal}{Phys.
  Rev. B} \textbf{\bibinfo{volume}{74}}, \bibinfo{pages}{104117}
  (\bibinfo{year}{2006}).

\bibitem[{\citenamefont{Likos}(2006{\natexlab{b}})}]{LikosNat}
\bibinfo{author}{\bibfnamefont{C.~N.} \bibnamefont{Likos}},
  \bibinfo{journal}{Nature} \textbf{\bibinfo{volume}{440}},
  \bibinfo{pages}{433} (\bibinfo{year}{2006}{\natexlab{b}}).

\bibitem[{\citenamefont{Pamies et~al.}(2009)\citenamefont{Pamies, Cacciuto, and
  Frenkel}}]{pephertz}
\bibinfo{author}{\bibfnamefont{J.}~\bibnamefont{Pamies}},
  \bibinfo{author}{\bibfnamefont{A.}~\bibnamefont{Cacciuto}}, \bibnamefont{and}
  \bibinfo{author}{\bibfnamefont{D.}~\bibnamefont{Frenkel}},
  \bibinfo{journal}{J. Chem. Phys.} \textbf{\bibinfo{volume}{131}},
  \bibinfo{pages}{044514} (\bibinfo{year}{2009}).

\bibitem[{\citenamefont{Mladek et~al.}(2008)\citenamefont{Mladek, Kahl, and
  Likos}}]{mladek}
\bibinfo{author}{\bibfnamefont{B.~M.} \bibnamefont{Mladek}},
  \bibinfo{author}{\bibfnamefont{H.}~\bibnamefont{Kahl}}, \bibnamefont{and}
  \bibinfo{author}{\bibfnamefont{C.~N.} \bibnamefont{Likos}},
  \bibinfo{journal}{Phys. Rev. Lett.} \textbf{\bibinfo{volume}{100}},
  \bibinfo{pages}{028301} (\bibinfo{year}{2008}).

\bibitem[{\citenamefont{Likos et~al.}(2001)\citenamefont{Likos, Hoffmann, and
  L\"owen}}]{likos}
\bibinfo{author}{\bibfnamefont{C.~N.} \bibnamefont{Likos}},
  \bibinfo{author}{\bibfnamefont{N.}~\bibnamefont{Hoffmann}}, \bibnamefont{and}
  \bibinfo{author}{\bibfnamefont{H.}~\bibnamefont{L\"owen}},
  \bibinfo{journal}{Phys. Rev. E} \textbf{\bibinfo{volume}{63}},
  \bibinfo{pages}{031206} (\bibinfo{year}{2001}).

\bibitem[{\citenamefont{Coluzza et~al.}(2011)\citenamefont{Coluzza, Capone, and
  Hansen}}]{capone22}
\bibinfo{author}{\bibfnamefont{I.}~\bibnamefont{Coluzza}},
  \bibinfo{author}{\bibfnamefont{B.}~\bibnamefont{Capone}}, \bibnamefont{and}
  \bibinfo{author}{\bibfnamefont{J.~P.} \bibnamefont{Hansen}},
  \bibinfo{journal}{Soft Matter} \textbf{\bibinfo{volume}{7}},
  \bibinfo{pages}{5255} (\bibinfo{year}{2011}).

\bibitem[{\citenamefont{Bolhuis et~al.}(2001)\citenamefont{Bolhuis, Louis, and
  Hansen}}]{Louis22}
\bibinfo{author}{\bibfnamefont{P.~G.} \bibnamefont{Bolhuis}},
  \bibinfo{author}{\bibfnamefont{A.~A.} \bibnamefont{Louis}}, \bibnamefont{and}
  \bibinfo{author}{\bibfnamefont{J.~P.} \bibnamefont{Hansen}},
  \bibinfo{journal}{Phys. Rev. E.} \textbf{\bibinfo{volume}{64}},
  \bibinfo{pages}{021801} (\bibinfo{year}{2001}).

\bibitem[{\citenamefont{Clark and Guenza}(2010)}]{Marina}
\bibinfo{author}{\bibfnamefont{A.~J.} \bibnamefont{Clark}} \bibnamefont{and}
  \bibinfo{author}{\bibfnamefont{M.~G.} \bibnamefont{Guenza}},
  \bibinfo{journal}{J. Chem. Phys.} \textbf{\bibinfo{volume}{132}},
  \bibinfo{pages}{044902} (\bibinfo{year}{2010}).

\bibitem[{\citenamefont{Jun and Arnold}(2007)}]{Jun1}
\bibinfo{author}{\bibfnamefont{S.}~\bibnamefont{Jun}} \bibnamefont{and}
  \bibinfo{author}{\bibfnamefont{A.}~\bibnamefont{Arnold}},
  \bibinfo{journal}{Phys. Rev. Lett.} \textbf{\bibinfo{volume}{98}},
  \bibinfo{pages}{128303} (\bibinfo{year}{2007}).

\bibitem[{\citenamefont{Grosberg and Khokhlov}(1994)}]{grossberg}
\bibinfo{author}{\bibfnamefont{A.~Y.} \bibnamefont{Grosberg}} \bibnamefont{and}
  \bibinfo{author}{\bibfnamefont{A.~R.} \bibnamefont{Khokhlov}},
  \emph{\bibinfo{title}{Statistical Physics of Macromolecules}}
  (\bibinfo{publisher}{American Institute of Physics}, \bibinfo{address}{New
  York}, \bibinfo{year}{1994}).

\bibitem[{\citenamefont{Miller and Cacciuto}(2009)}]{cacciuto}
\bibinfo{author}{\bibfnamefont{W.~L.} \bibnamefont{Miller}} \bibnamefont{and}
  \bibinfo{author}{\bibfnamefont{A.}~\bibnamefont{Cacciuto}},
  \bibinfo{journal}{Phys. Rev. E} \textbf{\bibinfo{volume}{80}},
  \bibinfo{pages}{021404} (\bibinfo{year}{2009}).

\bibitem[{\citenamefont{Sakaue and Rapha${\rm \ddot{e}}$l}(2006)}]{sakaue}
\bibinfo{author}{\bibfnamefont{T.}~\bibnamefont{Sakaue}} \bibnamefont{and}
  \bibinfo{author}{\bibfnamefont{E.}~\bibnamefont{Rapha${\rm \ddot{e}}$l}},
  \bibinfo{journal}{Macromolecules} \textbf{\bibinfo{volume}{39}},
  \bibinfo{pages}{2621} (\bibinfo{year}{2006}).

\bibitem[{\citenamefont{Frenkel and Smit}(2002)}]{FrenkelBook}
\bibinfo{author}{\bibfnamefont{D.}~\bibnamefont{Frenkel}} \bibnamefont{and}
  \bibinfo{author}{\bibfnamefont{B.}~\bibnamefont{Smit}},
  \emph{\bibinfo{title}{Understanding Molecular Simulation: From Algorithms to
  Applications}} (\bibinfo{publisher}{Academic Press},
  \bibinfo{address}{London}, \bibinfo{year}{2002}).

\bibitem[{\citenamefont{Rudnick and Gaspari}(1986)}]{rudnick}
\bibinfo{author}{\bibfnamefont{J.}~\bibnamefont{Rudnick}} \bibnamefont{and}
  \bibinfo{author}{\bibfnamefont{G.}~\bibnamefont{Gaspari}},
  \bibinfo{journal}{J. Phys. A: Math. Gen.} \textbf{\bibinfo{volume}{19}},
  \bibinfo{pages}{L191} (\bibinfo{year}{1986}).

\bibitem[{\citenamefont{Jagodzinski et~al.}(1992)\citenamefont{Jagodzinski,
  Eisenriegler, and Kremer}}]{Jado}
\bibinfo{author}{\bibfnamefont{O.}~\bibnamefont{Jagodzinski}},
  \bibinfo{author}{\bibfnamefont{E.}~\bibnamefont{Eisenriegler}},
  \bibnamefont{and} \bibinfo{author}{\bibfnamefont{K.}~\bibnamefont{Kremer}},
  \bibinfo{journal}{J. Phys. I (France)} \textbf{\bibinfo{volume}{2}},
  \bibinfo{pages}{2243} (\bibinfo{year}{1992}).

\bibitem[{\citenamefont{Cannon et~al.}(1991)\citenamefont{Cannon, Aronovitz,
  and Goldbart}}]{Cannon}
\bibinfo{author}{\bibfnamefont{J.~W.} \bibnamefont{Cannon}},
  \bibinfo{author}{\bibfnamefont{J.~A.} \bibnamefont{Aronovitz}},
  \bibnamefont{and} \bibinfo{author}{\bibfnamefont{P.}~\bibnamefont{Goldbart}},
  \bibinfo{journal}{J. Phys. I (France)} \textbf{\bibinfo{volume}{1}},
  \bibinfo{pages}{629} (\bibinfo{year}{1991}).

\bibitem[{\citenamefont{Daoud and Cotton}(1982)}]{cotton}
\bibinfo{author}{\bibfnamefont{M.}~\bibnamefont{Daoud}} \bibnamefont{and}
  \bibinfo{author}{\bibfnamefont{J.}~\bibnamefont{Cotton}},
  \bibinfo{journal}{J. Phys. (Paris)} \textbf{\bibinfo{volume}{43}},
  \bibinfo{pages}{531} (\bibinfo{year}{1982}).

\end{thebibliography}

\end{document}